\documentclass[10pt,letterpaper]{article}
\usepackage{opex3}
\usepackage{cite}

\newcommand{\text}[1]{\mathrm{#1}}
\begin{document}

\title{Spatio-temporal characterization of mid-infrared laser pulses\\
with spatially encoded spectral shearing interferometry}

\author{T. Witting,$^*$ S.J. Weber, J.W.G. Tisch, and J.P. Marangos}
\address{Blackett Laboratory, Imperial College, London SW7 2AZ,UK}

\email{$^*$ t.witting@imperial.ac.uk} 

\homepage{http://www.attosecond.org} 


\begin{abstract}
We report on the spatially resolved full amplitude and phase characterization of mid-infrared high intensity laser pulses generated in a three stage OPA. We use a spatially-encoded arrangement (SEA-)SPIDER with spectral filters for ancilla generation for spatially resolved characterization. Using five interchangeable filter sets we are able to characterize pulses from 1 to 2\,$\mu{}$m with one single device with minimal adjustments.
\end{abstract}


\bibliographystyle{osajnl}


\section{Introduction}
A significant part of future development in attosecond science~\cite{krausz_attosecond_2009} and especially molecular high harmonic generation spectroscopy~\cite{marangos_dynamic_2007} is likely to require the use of intense and ultrashort infrared laser pulses in the mid-infrared (mid-IR) spectral range. Recently mid-IR pulses have been employed to observe the Cooper minimum in Krypton~\cite{shiner_observation_2012}.
Secondary sources based on Ti:Sapphire laser pumped OPA in nonlinear crystals are readily available as commercial systems. Few-cycle mid-IR sources have been demonstrated~\cite{cerullo_ultrafast_2003,schmidt_compression_2010,cerullo_few-optical-cycle_2011}, and much progress has been made in the development of optical parametric chirped pulse amplification (OPCPA) systems~\cite{chalus_mid-ir_2009}.
Despite rapid progress on mid-IR laser sources, their temporal characterization is still a challenge and underdeveloped compared to the state of the art for pulse characterization at near-IR wavelengths. An excellent overview over the characterization of ultrashort pulses can be found in~\cite{walmsley_characterization_2009}. We briefly review mid-IR related pulse characterization efforts. Second harmonic generation frequency resolved optical gating (SHG-FROG) has been used to measure the pulse duration of 5\,$\mu$m free electron laser pulses~\cite{richman_temporal_1997}. Mid-IR pulses at 3.2\,$\mu$m from an OPCPA system have been characterized with SHG-FROG~\cite{bates_ultrashort_2010}. Cross-correlation FROG has been employed for the measurement of 4.0\,$\mu$m pulses in~\cite{reid_amplitude_2000}. Ventalon et al. employed a time-domain homodyne optical technique for spectral phase interferometry for direct electric-field
reconstruction (HOT SPIDER) approach for the characterization of pulses at 9\,$\mu$m\cite{ventalon_generation_2006}.

To our knowledge, all thus far demonstrated mid-IR pulse characterization methods are multi-shot and offer no spatial resolution. Almost all mid-IR sources are based on nonlinear optical processes, 
and spatio-temporal effects can play a role in the generation stage. Therefore the development of a wavelength flexible and spatially resolving pulse characterization tool is much needed. In this paper we report the spatially resolved characterization of up to 1.8\,mJ, 26 to 70\,fs mid-IR pulses ranging in wavelength from 1.1 up to 2.2\,$\mu$m.


\section{Mid-IR SPIDER apparatus}

We characterize mid-IR pulses with a spatially encoded arrangement for direct electric field reconstruction by spectral shearing interferometry (SEA-SPIDER) previously used to characterize near-IR pulses centered at 800\,nm~\cite{kosik_interferometric_2005}. SPIDER has the feature that the signal wavelength can be freely chosen by selecting an appropriate ancilla wavelength. SEA-SPIDER offers significant advantages over the traditional SPIDER device and is especially suited for few-cycle pulse characterization~\cite{witting_characterization_2011}. The experimental design allows easy tuning of the ancilla frequencies without changing any other device parameters. SEA-SPIDER is a zero-additional phase measurement and also benefits from less demanding spectrometer resolution requirements. For a pulse with field
$E(y,t)=\int_{-\infty}^{+\infty} A(y,\omega)\,\exp{[i\phi(y,\omega)-i\omega\,t]} dt$, where $A(y,\omega)$ and $\phi(y,\omega)$ are the spectral amplitude and phase, respectively, and $y$ is the transverse spatial coordinate (here the vertical direction), the SEA-SPIDER trace is given by
\begin{eqnarray}
  \label{eq:seaspidertrace}
\nonumber
  S(y, \omega) &=& |E(y,\omega)|^2 +  |E(y,\omega-\Omega)|^2 +\\
\nonumber &&  2|E(y,\omega)|\,|E(y,\omega-\Omega)| \times\\
&&  \cos{[\phi(y,\omega)-\phi(y,\omega-\Omega)+\Delta k y]},
\end{eqnarray}
where $\Omega$ is the spectral shear and $\Delta k$ is the wavevector difference arising from the angle between the two signal beams.
From $S(y, \omega)$ the spatially resolved spectrum $A(y,\omega)$, and the phase $\phi(y,\omega)$ can be reconstructed with two-dimensional Fourier filtering, resulting in reconstruction of the temporal field $E(y,t)$, thus enabling single-shot spatially resolved pulse characterization. 

\subsection{Experimental Setup}

\begin{figure}[h!]
  \centering
  \includegraphics[width=0.7\columnwidth]{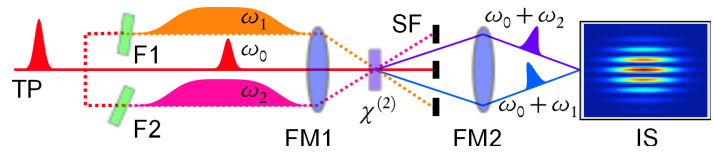}
  \caption{Concept of the mid-IR SEA-F-SPIDER apparatus. Please refer to the text for a detailed explanation.}
  \label{fig:setup}
\end{figure}

A sketch of the experimental setup is shown in Fig.~\ref{fig:setup}.
The unknown testpulse is split into three parts. A small fraction (5\%) is split off the test-pulse and is guided to the crystal for upconversion with the ancillae beams using only reflective optics. The remaining part is split into equal halfs to generate two ancilla beams. Each ancilla beam is filtered by a narrowband transmission filter (F1, and F2). The two ancilla beams and the test-pulse beam are focused with a spherical mirror (FM1) to undergo Type-I sum-frequency mixing in a 50\,$\mu$m (100\,$\mu$m) thick BBO crystal for the 1.4 and the 2.2\,$\mu$m pulses, respectively. We choose Type-I phasematching over the in SPIDER commonly used Type-II variant to avoid using a waveplate in the ancilla beam. The phasematching bandwidth (see red line in Fig.~\ref{fig:ancilla}(c)) is sufficiently broad to cover our desired test pulse range with one single crystal. After blocking the unwanted fundamental beams by spatial filter SF the two SPIDER signal beams are re-imaged with a focusing mirror (FM2) onto the entrance slit of a homemade imaging spectrometer~\cite{austin_broadband_2009}. The signal is dispersed by a 150\,l/mm Al coated grating (Edmund Optics). The SPIDER signal is observed in the first grating order and the fundamental signal components are blocked out in the image plane of the spectrometer to avoid the use of color glass bandpass filters. The grating can be tuned to center the spectrometer on the SPIDER signal for different test pulses. The SPIDER signal $S(y, \omega)$  is recorded on a 640$\times$480\,pixel, 14\,bit silicon based CCD camera (AVT GmbH).

\subsection{Ancilla generation and choice of signal wavelength}
\label{sec:ancilla}

\begin{figure}[htbp]
  \centering
  \includegraphics[width=0.7\columnwidth]{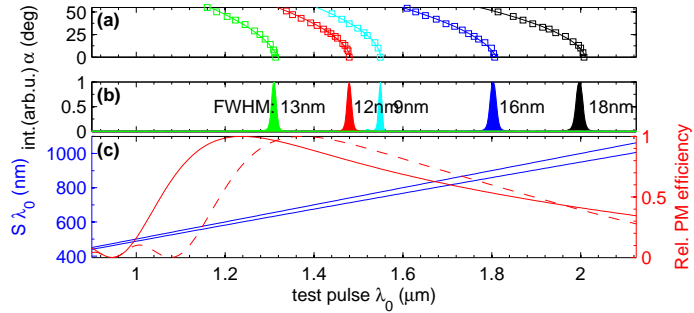}
\caption{Ancilla filters tuning range: (a) transmitted beam center wavelength as function of filter angle, (b) Transmission spectra for normal incidence. (c) SPIDER signal wavelength (range indicated by blue lines); phasematching efficiency (50\,$\mu$m BBO red solid line, 100\,$\mu$m BBO red dashed line). Details in text.}	
  \label{fig:ancilla}
\end{figure}

We employ the SEA-SPIDER with direct spectral filtering for ancilla preparation (SEA-F-SPIDER)~\cite{witting_improved_2009} to increase the accuracy in determining the two important calibration parameters spectral shear $\Omega = \omega_2-\omega_1$ and upconversion frequency $\omega_{\text{up}}=\omega_1$ (definitions in Fig.~\ref{fig:setup}). 
The choice of ancilla wavelength is entirely free in spectral shearing interferometry~\cite{walmsley_characterization_2009}. For example one can use ancilla beams around 800\,nm (derived from the unused TOPAS pump light) to upconvert mid-IR spectra to the visible. The SPIDER signal will by spectrally centered around $\omega_0 + \omega_{\text{up}}$. This means for a given test pulse center frequency $\omega_0$ the choice in $\omega_{\text{up}}$ allows to move the signal to a wavelength where the used detector has a good quantum efficiency. For example a 5\,$\mu$m test pulse could be characterized using an ancilla beam centered at 800\,nm to generate a SPIDER signal at 690\,nm, which lies in a very responsive area of silicon based CCD detectors. However, the use of an external ancilla pulse would require an external delay-line to temporally overlap the unknown mid-IR pulse with the 800\,nm ancilla pulses. For the current work we had only pulses available up to 2.0\,$\mu$m and therefore a short wavelength ancilla is not necessary. We used part of the unknown pulse's spectrum to derive the ancilla beams. The SPIDER signal is centered in the near-IR (range indicated by blue lines in Fig.~\ref{fig:ancilla} (c)), where it can be detected with a standard silicon based CCD. In a SPIDER with glass block stretcher or grating compressor, the stretching optics would have to be adapted for each different wavelength. The main drawback is that most glasses can have very low dispersion in the mid-IR spectral range. A grating compressor would provide adequate dispersion, but is more complex to setup and has to be retuned for each test pulse wavelength.

For those reasons, and the enhanced precision the filter-SPIDER concept offers, the SEA-F-SPIDER with direct spectral filtering is an ideal device for wavelength tunable use in the mid-IR. Filters with narrow transmission are readily available. Rotation of the filter angle $\alpha$ with respect to the input beam allows for tuning of the transmission frequency and hence setting of the upconversion frequency $\omega_{\text{up}}$, and the spectral shear $\Omega$. Different easily interchangeable filter sets (in order of wavelength: Thorlabs FB1310-12, Thorlabs, FB1480-12, Semrock, 1550/3, Knight Optical, 1800-T-FKH, Knight Optical, 2000-17-T-DRC) as shown in Figs.~\ref{fig:ancilla}(a) and \ref{fig:ancilla}(b) allow quick, straightforward adaptation of our device to the test pulse signal wavelength. Only small re-adjustments of the ancilla to test-pulse time-delays are necessary after a filter change to account for differences in thickness between different filter sets. Figure~\ref{fig:ancilla}(a) shows the tuning range of the ancilla filters. We can generate ancilla beams from 1.2 up to 2.0\,$\mu$m, resulting in SPIDER signals centered from 450 to 1000\,nm (blue lines in Fig.~\ref{fig:ancilla}). The shear will be set by using two filters at slightly different angles. Very large shears of up to 35\,mrad/fs are easily obtainable with a 25\,deg filter rotation $\alpha$. It should be noted that the filter only needs to cover a spectral region with sufficient intensity of the unknown pulse. Hence, broadband pulses with spectral content well above 2.0\,$\mu$m can be characterized in our device as long as there is sufficient intensity around 2.0\,$\mu$m. For longer testpulse wavelengths the silicon based CCD could be reaplced or an external ancilla beam at 800\,nm could be used.

\subsection{Spatio-temporal Reconstruction}

SEA-SPIDER offers not only single shot capable pulse characterization, but also the acquired spatial information can be exploited. The spectral phase and the pulse spectrum is spatially resolved. In this work the spatially resolved fundamental spectrum is extracted as absolute value from the AC term (lines 2 and 3 in Eq.(\ref{eq:seaspidertrace})) recovered via Fourier filtering of the SEA-F-SPIDER trace. After correction for spectral response of the apparatus and subtraction of the upconversion frequency the spatially resovled spectral amplitude $|A(y,\omega)|$ is recovered, which can be used with the spatially resolved spectral phase $\phi(y,\omega)$ to perform a space-time reconstruction of the unknown pulse. For a full spatio-temporal pulse reconstruction the spectral as well as the spatial phase has to be measured. This requires simultaneous spectral, and also spatial shearing interferometry and has been demonstrated for 800\,nm pulses~\cite{dorrer_direct_2002,dorrer_spatio-temporal_2002}. Spatial shearing would require the implementation of an additional spatial shearing interferometer into the SEA-SPIDER device. 
\begin{figure}[h!]
  \centering
  \includegraphics[width=0.9\columnwidth]{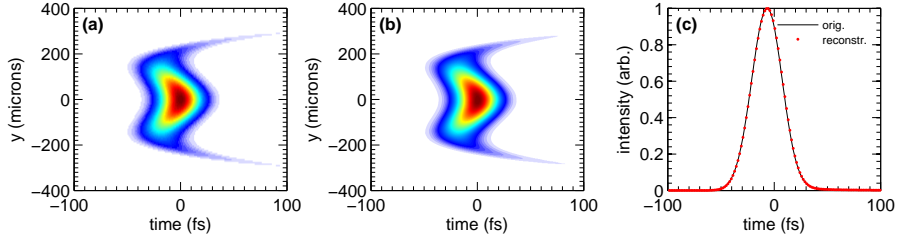}
\caption{Spatio-temporal reconstruction with SEA-SPIDER. (a) simulated spatio-temporal intensity $|E(y,t)|^2$, (b) SEA-F-SPIDER reconstruction, (c) 1D marginals. For details please refer to the text.}	
  \label{fig:spatio-temporal}
\end{figure}
Even without this modification a SEA-SPIDER offers spatial resolution along one spatial slice of the beam. With the spatially filtered ancilla beams acting as a spatial phase reference. Because no spatial shear is introduced the spatial information is incomplete and pulse front tilt cannot be measured. However, spatio-temporal couplings of higher orders can be recovered from a SEA-SPIDER trace~\cite{wyatt_accuracy_2011}. We illustrate this with numerical simulations for space-time coupled pulses. 
Figure~\ref{fig:spatio-temporal} shows the simulated space-time field $|E(y,t)|^2$ of a spatio-temporally coupled pulse with space-time coupling of negative second and positive third order. The SEA-F-SPIDER reconstruction shown in (b) recovers the space-time structure correctly. The temporal marginals of original and reconstructed pulses are shown in Fig.~\ref{fig:spatio-temporal}(c) to show the excellent agreement.


\section{Pulse characterization experimental results}

We generate ultrashort mid-IR pulses with energies of up to 1.8\,mJ, 26 to 70\,fs ranging in wavelength from 1.1 up to 2.2\,$\mu$m  in a three stage optical parametric amplifier (OPA) (HE-TOPAS, Light Conversion). The OPA signal pulse is tunable from 1.1 to 1.55\,$\mu$m. The idler is tunable from 1.7 to 2.2\,$\mu$m. The OPA is pumped by 8\,mJ, 28\,fs pulses at 1\,kHz repetition rate from a chirped pulse amplification system (CPA), (RedDragon, KMlabs).

\begin{figure}[h!]
  \centering
  \includegraphics[width=0.5\columnwidth]{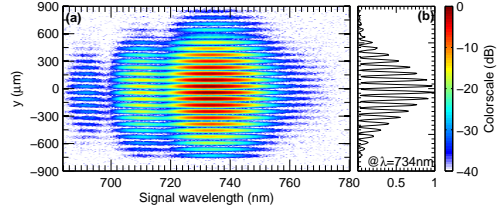}
  \caption{SEA-F-SPIDER data trace for a 1450\,nm test-pulse. 40\,dB colour scale. 40 laser shots integration time}
  \label{fig:trace1450nm}
\end{figure}

\begin{figure}[h!]
  \centering
  \includegraphics[width=\textwidth]{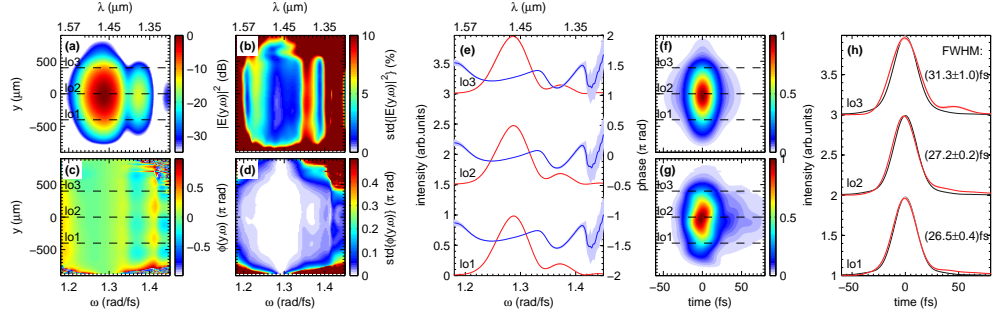}
  \caption{Spatially resolved pulse reconstruction for a 1.4 $\mu$m pulse. (a) to (e) spectral domain, (f) to (h) 	temporal domain.
	(a) Mean of 25 spectra $|E(y,\omega)|^2$ (40\,dB colour scale).
	(b) Relative standard deviation of $|E(y,\omega)|^2$ in percent.
	(c) Mean of spectral phase $\langle\varphi(y, \omega)\rangle$.
	(d) Standard deviation of phase $\sigma[\varphi(y, \omega)]$.
	(e) Lineouts of the spectral intensity and phase at the $y_i$ positions indicated in (a) and (c) by the black dashed lines. The shaded area represents the $\pm 1 \sigma$ interval. 
	(f) Fourier-limited temporal intensity $|E(y,t)|^2$ (lin. colour scale).
	(g) Temporal intensity $|E(y,t)|^2$ (lin. colour scale, same $y$-axis as (a) to (d)).
	(h) $|E(y_i,t)|^2$ for the spatial positions $y_i$ indicated by the dashed black horizontal lines in (a), (c), and (f)/(g). Black is the Fourier-limited pulse. The red curve is the actual pulse.
  \label{fig:data1450nm}}
\end{figure}

In principle, single laser shot recording of the complete data trace is possible with SEA-F-SPIDER, but we are using a standard silicon based CCD, so the signal strength is relatively low especially for the test-pulse wavelengths above 1.8\,$\mu$m, for which the SPIDER signal is centered in the poor quantum efficiency region of the detector beyond 900\,nm. Therefore we use integration times of typically 20\,ms. We record the ancilla spectra individually by blocking the testpulse and removing the spatial filter SF (see Fig.~\ref{fig:setup}) to directly measure the upconversion frequency $\omega_{\text{up}}$, and the spectral shear $\Omega$ before each data acquisition.
Figure~\ref{fig:trace1450nm}(a) shows a SPIDERgram for 1.4\,$\mu$m test pulses. The signal is shown on 40\,dB color scale.  A spatial lineout showing the spatial fringes due to the spatial beam tilt $\Delta k$ is shown in Fig.~\ref{fig:trace1450nm}(b).

\begin{figure}
  \centering
  \includegraphics[width=\textwidth]{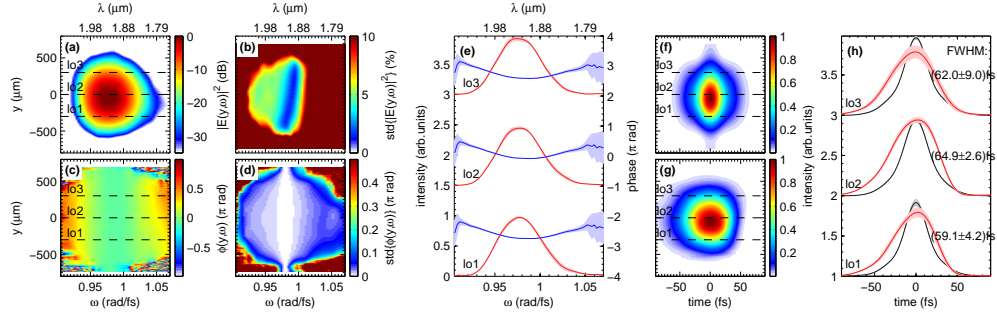}
  \caption{Spatially resolved pulse reconstruction for a 1.9\,$\mu$m pulse. For a description of the subplots see the caption of Fig.~\ref{fig:data1450nm}.
	\label{fig:data1900nm}}
\end{figure}

Figure~\ref{fig:data1450nm} shows the pulse characterization results for a TOPAS signal pulse at 1.45\,$\mu$m central wavelength. The reconstructed spatially resolved spectral intensity $|E(y,\omega)|^2$ of the pulse is shown in Fig.~\ref{fig:data1450nm}(a). Note the high signal to noise boost resulting from the two-dimensional Fourier filtering routine in comparison to the corresponding data in Fig.~\ref{fig:trace1450nm}(a).
The reconstructed spectral phase is shown in (c). The phase looks to be spatially homogenous apart from a few islands in the nodes of the spectral intensity. The fluctuations in spectral intensity retrieved from 25 consecutive acquisitions of SPIDER traces are shown in (b). As expected the intensity fluctuations are higher in the lower intensity parts of the spectrum. We can also deduce from (a) that the beam is virtually free of spatial chirp. Figure~\ref{fig:data1450nm}(d) shows the standard deviation of the reconstructed spectral phase $\sigma[\varphi(y, \omega)]$. It is remarkable that the phase is recovered well even in the low intensity parts of the spectrum. This is a feature of the superior noise rejection of the 2D Fourier filtering~\cite{witting_ultrashort_2009}. A few one-dimensional lineouts at different spatial positions of the spectral intensity and phase are shown in Fig.~\ref{fig:data1450nm}(e). The blue shaded area marks the $\pm1$ standard deviation intervals derived from the 25 successively recorded data traces. Figures~\ref{fig:data1450nm}(f) to \ref{fig:data1450nm}(h) show the spatially resolved temporal pulse reconstruction. the Fourier transform limited pulse is shown in (f). The actual pulse is shown in (g). A few spatial lineouts are shown in (h). The pulse duration in the beam center is $27.2\pm0.2$\,fs. The pulse is spatially homogenous and shows deviations only far off center in the shoulders as shown by lineout 3 and 1 at $y_i=300$, and $-300\,\mu$m respectively. Figure~\ref{fig:data1900nm} shows the spatially resolved reconstruction of the TOPAS idler at 1.9\,$\mu$m center wavelength. The pulse is slightly chirped and we measure a pulse duration of $64.9\pm2.6$\,fs in the beam center. The pulse is spatially very homogenous in the higher intensity parts with deviations only in the low energy wings.

\section{Conclusion}
In conclusion we have demonstrated the spatially resolved characterization of ultrashort mid-IR pulses for the first time. Our device can reconstruct the pulse field (spectrum and phase) from a data trace with all information contained in principle in a single shot data trace. It is flexible in terms of test pulse wavelength and allows the extraction of spatial information and at the same time the recording of statistics of the laser source fluctuations. We envisage such spatially revolved characterization to be vital for ultrashort and few-cycle mid-IR pulses and it will greatly aid in the compression and optimization of the source performance.

\section*{Acknowledgments}
This research was supported in part by EPSRC programme grant EP/I032517/1, and EPSRC translation grant
EP/F034601/1. 
T.W. and S.J.W. contributed equally to the work presented.

\end{document}